\documentclass[pre,aps,floatfix,amsmath,amssymb,twocolumn,footinbib,10pt]{revtex4-1}
\usepackage{bm,color,float,dcolumn,amsmath,amssymb,amsfonts,amsthm,nicefrac,graphicx,subfigure,dsfont}
\usepackage{dsfont,afterpage}
\definecolor{darkred}{rgb}{0.8,0.0,0.0}
\definecolor{darkblue}{rgb}{0.0,0.0,0.8}
\usepackage[colorlinks,urlcolor=darkred,linkcolor=red,anchorcolor=darkred,citecolor=red]{hyperref}
\usepackage[english]{babel}
\newcommand{\Mod}[1] {\ ( \text{mod}\ #1)}
\newcommand{\Bra}[1]{\left| #1 \right\rangle}

\begin{document}

\title{Parafermion chain with $2\pi/k$ Floquet edge modes}
\author{G.~J.~Sreejith,  Achilleas Lazarides and Roderich Moessner}
\affiliation{Max Planck Institute f\"{u}r Physik komplexer Systeme, 01187 Dresden, Germany}
\date{\today}
\begin{abstract}
We study parafermion chains with $\mathbb{Z}_k$ symmetry subject to a periodic binary drive. We focus on the case $k=3$. We find that the chains support different Floquet edge modes at nontrivial quasienergies,  distinct from those for the static system. We map out the corresponding phase diagram by a combination of analytics and numerics, and provide the location of $2\pi/3$ modes in parameter space. We also show  that the modes are robust to weak disorder. While the previously studied $\mathbb{Z}_2$-invariant Majorana systems posesses a transparent weakly interacting case where the existence of a $\pi$-Majorana mode is manifest,  our intrinsically strongly interacting generalization demonstrates that the existence of such a limit is not necessary. 
\end{abstract}

\maketitle
\section{Introduction}

The advent of experimental systems exhibiting long quantum mechanical coherence times has generated 
much interest in the question of how macroscopically observable 
behavior emerges from the underlying many-body time evolution. In particular, 
the possibility of driving coherent many-body
systems out of equilibrium has opened a new window on the fundamental question of how to describe
systems for which the concepts of equilibrium thermodynamics are a priori inapplicable
\cite{Polkovnikov:2011iu,DAlessio:2015tw}. 

For periodically driven systems, there has been considerable progress in understanding how the roles 
of equilibration, thermalization and ordering, 
need to be modified in order to generalize the notions from thermodynamic equilibrium to this out of equilibrium situation.~\cite{Lazarides:2014cl,Russomanno:2012bf,DAlessio:2014fg,Lazarides:2014ie,Ponte:2015hm} 
While the notion of temperature is lost in the absence of a time-translation invariant Hamiltonian, three
distinct thermodynamic ensembles have been identified: a generic 
Floquet-ETH ensemble, corresponding to an `infinite
temperature' ensemble of the undriven Hamiltonian; a periodic Gibbs ensemble generalising the 
concept of the generalised Gibbs ensemble of integrable undriven systems; and finally, a Floquet-MBL
ensemble, where ergodicity breaking due to many-body localisation occurs~\cite{Lazarides:2015jd,Ponte:2015dc}

While the Floquet-ETH ensemble does not support any notion of ordering on account of its entirely featureless `infinite temperature' correlations, a recently proposed notion of ordering in the presence of MBL does generalise to the Floquet-MBL regime, namely that of eigenstate ordering~\cite{Huse:2013bw,Pekker:2014bj}. The notion of eigenstate order appeals to the idea of symmetry-breaking manifesting itself in thermodynamic-limit degeneracies of eigenstates. Partners in such quasi-degenerate multiplets of many-body eigenstates can exhibit   finite expectation value of an Edwards-Anderson-like order parameter, or indeed topological forms of order, not only in the ground state but also in the excited state spectrum. 
 
The generalisation of this set of ideas to the Floquet realm proceeds via the recognition that Floquet systems also support the notion of eigenstates with a quasi-energy (defined modulo the energy corresponding to the driving frequency, $\hbar\omega$)\cite{Khemani:2015wn};  this is provided by the eigensystem of the stroboscopic time evolution operator over a driving period $T=2 \pi/\omega$. Besides extending the notion of eigenstate order to the Floquet realm, this has allowed the definition of new types of order, distinguishable only in the time-dependence of the order parameter, which have no correspondence in the undriven setting~\cite{Khemani:2015wn}. 

Investigations of this new type of order, e.g.\ from the point of view of its universal properties~\cite{Gannot2015}, is only at its beginning. Here, we present an analysis of a driven $\mathbb{Z}_n$ symmetric clock model and its edge mode properties \cite{Fendley2012,Jermyn2014,Ostlund1981,Huse1981,Howes1983,Zhuang2015}. We focus on the $\mathbb{Z}_3$-symmetric clock-model, which is equivalent to a $\mathbb{Z}_3$ parafermion chain under a Jordan Wigner transformation.

The motivation for this is threefold. Firstly, it is to provide further examples of the new Floquet ordering phenomena beyond the very simplest and so far only studied case of $\pi$ Floquet-Ising order\cite{Khemani:2015wn,Gannot2015,Jiang2011,Benito:2014bd,Thakurathi2013}, in order to assist and test more general schemes for classifying Floquet-MBL ordering physics. Secondly, unlike  the driven Ising models previously studied, this model has no non-interacting limit. This poses both questions of principle, about the role presence or absence  of such a limit plays, and raises the question in practice how to locate novel ordered phases and illuminate their properties in the absence of the non-interacting solution as a guide. Thirdly, the model we consider is of autonomous interest as an implementation of an anyonic chain, with the aim of realizing platforms for topological quantum computation--edge modes of parafermionic chains  provide a starting point for achieving approximations to general qubit rotations. In light of the previous studies indicating presence of edge modes in a driven Majorana chain, it is natural to consider the edge properties of a periodically driven parafermionic chain. 
 
We find that there does indeed exist  rich edge mode physics in the parafermion chains. In contrast to the previously studied $\mathbb{Z}_2$ case of a driven Majorana chain, in which the algebra of the edge operator $\Psi^2=1$ forces its quasi-energy to be $0$ or $\pi$, a parafermionic edge mode satisfying $\Psi^3=1$ can have an energy of $0$ or $\pm\frac{2\pi}{3}$. The presence of a $\pm\frac{2\pi}{3}$ mode implies that $\Psi$ and $\Psi^{2}$ add $2\pi/3$ and $-2\pi/3$ (equivalent to $4\pi/3$) to the quasi-energy of a state, and $\Psi^3$ acts trivially on any state. The presence of a zero mode implies $\Psi,\Psi^2$ map states to new ones with the same quasienergy. A general $\mathbb{Z}_n$ symmetric system can instead have modes of quasienergies $2\pi/k$ where $k$ is any factor of $n$. For the case of $n=3$, we determine the locations of the parameter regimes supporting these $0$ and $\pm\frac{2\pi}{3}$ modes, 
provide diagnostics confirming their presence and explore the phase diagram using the spectral function of the edge clock-operators in a finite sized open chain numerically. We also address their robustness by considering the evolution of the diagnostics when
disorder is added.

The article is set out as follows. In Sec.~\ref{Sec:Znmodel}, we describe the periodically driven $\mathbb{Z}_n$ symmetric system. The protocol proposed is a generalization of the binary drive protocol previously shown to result in zero and $\pi$ Majorana edge modes in a transverse field Ising model.\cite{Jiang2011,Benito:2014bd,Thakurathi2013} We specifically focus on the $\mathbb{Z}_3$ case, describing the model in the static and the driven case. In Sec-\ref{General}, using general arguments, we show that the periodically driven system can have edge modes of quasi-energy $0$ or $\pm \frac{2\pi}{3}$ in contrast to the static system which only displays zero-energy modes. We deduce the parameter regimes where the $0$ and $\pm \frac{2\pi}{3}$ edge modes are expected. In Secs.~\ref{Numerics} we introduce the spectral function of an edge operator and use it to numerically map out a phase diagram. 

\section{Periodically driven $\mathbb{Z}_n$ chain}
\label{Sec:Znmodel}
The periodically driven $\mathbb{Z}_n$ clock chain can be described in terms of clock operators $\sigma_i$ associated with each lattice point of an open chain of length $L$. The clock at each site $i$ is associated with an $n$-dimensional Hilbert space, spanned by the $n$ orthonormal eigenvectors $\Bra{s}$, $s=0,2,\dots, n-1$ of $\sigma$ with eigenvalues $\omega^{s}$, where $\omega=\exp\left[\frac{2\pi\imath}{n}\right]$ is the $n^{\rm th}$ root of 1. An operator $\tau$ at each site shifts these eigenstates through $\tau\Bra{s}=\Bra{s+1\Mod{n}}$. The operators at different sites commute, while at a given site they satisfy the following algebra
\begin{equation}
\sigma^n=\tau^n=1;\;\sigma^\dagger = \sigma^{n-1};\;\tau^\dagger=\tau^{n-1};\;\sigma\tau=\omega\tau\sigma.\label{clockoperatoralgebra}
\end{equation}

\begin{figure}
\includegraphics[width=\columnwidth]{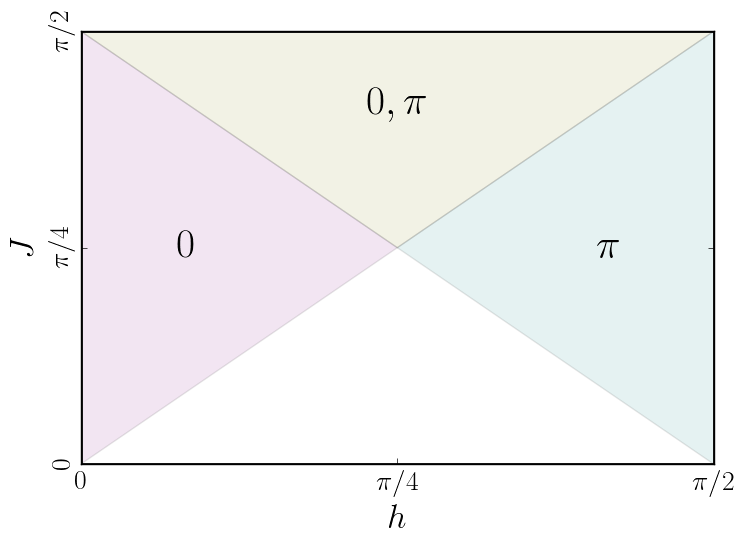}
\caption{Phase diagram of the periodically driven Majorana chain, showing parameter regimes where edge modes of quasi-energy $0, \pi$, or both, are present.}
\label{Fig:TFIM_periodic_phase}
\end{figure}

The driven system that we study represents a binary drive in which the system evolves with a Hamiltonian $H_J$ that stabilizes an ordered phase, and a Hamiltonian $H_h$ which stabilizes a disordered phase for the first and second halves of each time period. The state of the system $\Bra{\varphi_r}$ at time $t=rT$ evolves as    $\Bra{\varphi_{r+1}}=U\Bra{\varphi_{r}}$ where $U$ is given by 
\begin{eqnarray}
U(J,h) = \exp\left[- \imath  H_h\right] \exp\left[- \imath  H_J\right].
\end{eqnarray}
The Hamiltonians $H_h$ and $H_J$ are given by 
\begin{gather}
H_h = \sum_{i=1}^{L} \sum_{m=1}^{n-1} h_m \tau_i^{m}\label{drivenZn}\\
H_J = \sum_{i=1}^{L-1} \sum_{m=1}^{n-1} J_m [\sigma_i^\dagger \sigma_{i+1}]^m\nonumber
\end{gather}
The coupling constants $J_m,h_m\in \mathbb{C}$ satisfy $J_m=\bar{J}_{n-m}$ and $h_m=\bar{h}_{n-m}$, which enforces Hermiticity of the Hamiltonians. 

The eigenstates of the unitary operator $U$  are known as Floquet states, and their eigenvalues exp($i \epsilon_n$) yield quasi-energies $\epsilon_n$ defined  modulo $2 \pi$. Since $U$
commutes with the $\mathbb{Z}_n$ charge operator 
\begin{equation}
Q_n=\prod_{i=1}^{L} \tau_i\label{ZnCharge}
\end{equation}
the Floquet eigenstates are associated with a $\mathbb{Z}_n$ charge which takes values from $\{\omega^s\}_{s=1\to n}$

The simplest case, namely that of the $\mathbb{Z}_2$ symmetric system~\cite{Jiang2011,Thakurathi2013}, can be mapped, after a non-local Jordan Wigner transformation, to a system of Majorana fermions on a chain. In the language of the Majorana fermionic variables, the two Hamiltonians mentioned above stabilize phases with alternate dimerization patterns. The quadratic nature of the fermionic system allows a decomposition of the system into a set of non-interacting single-particle modes each carrying a fixed quasi-energy. A key feature of the model is the presence of modes localized at the edge of the system, and which can carry a quasi-energy of exactly $0$ or $\pi$. The phase diagram of the system (Fig-\ref{Fig:TFIM_periodic_phase}) \cite{Khemani:2015wn} has a detailed understanding based on a transfer matrix approach that relies critically on the non-interacting nature of the system \cite{Gannot2015}. 

In this work, we aim to understand case of the more general $\mathbb{Z}_n$ symmetric system. The key difference from the $\mathbb{Z}_2$ case is the strongly interacting nature preventing a decomposition into independent modes. We focus on the  $\mathbb{Z}_3$ system beginning with a summary of the results known in the static system.

\subsection{\texorpdfstring{$\mathbb{Z}_3$}{Z3} symmetric clock model}
\label{Sec:Z3model}
In addition to having a larger parameter space, the $\mathbb{Z}_3$ symmetric system is qualitatively different from the $\mathbb{Z}_2$ case due the absence of a non-interacting limit with independent modes. The presence of an independent edge mode in such a system is therefore not a priori a simple extension of the $\mathbb{Z}_2$ case. The properties of the static system will be shown in the latter sections to have bearings for domains in parameter space supporting the edge modes for the driven case.

\subsubsection{Static system}
\label{Sec:Z3modelStatic}
The static chiral $\mathbb{Z}_3$ clock model is described by a $\mathbb{Z}_3$ invariant nearest neighbor interaction Hamiltonian acting on a clock-chain \cite{Fendley2012,Ostlund1981,Huse1981}
\begin{equation}
H =  H_J+H_h= - J\sum_{i=1}^{L-1}  \sigma^\dagger_i \sigma_{i+1} - h \sum_{i=1}^L \tau_i + h.c.
\label{Eq:StaticHamiltonian}
\end{equation}
where $\sigma_i$ and $\tau_i$ act on the clock at the site $i$ of an open chain of length $L$, and $J,h \in \mathbb{C}$. The above Hamiltonian commutes with the $\mathbb{Z}_3$ charge operator $Q\equiv Q_3$ (Eq-\ref{ZnCharge}).

All eigenstates of the Hamiltonian are three-fold degenerate at $h=0$. For small finite $|\frac{h}{J}|$ in a finite system, it was argued in Ref-\onlinecite{Jermyn2014} that the three lowest energy eigenstates have an energy splitting that exponentially decays with system size,\cite{Motruk2013} whereas higher energy triplets are expected to have energy splittings that decay with a power law. However, for $\arg(J)$ outside a small finite neighborhood of $0\Mod{\pi/3}$, frustration in domain wall tunneling results in a regime of small $|\frac{h}{J}|$ within which all energy eigenstate triplets have energy splittings that exponentially decay with the system size. The most robust degeneracies occur at $\arg(J)=\frac{\pi}{6}\Mod{ \frac{\pi}{3}}$~\cite{Fendley2012,Jermyn2014}. These degeneracies disappear for $|\frac{h}{J}|>1$ \cite{Fendley2012,Zhuang2015}.

The clock-model can equivalently be described \cite{Pfeuty1970,Kitaev2001}  in a language where the local variables are parafermions $\chi$ and $\psi$, which are related to the above clock variables through a Jordan-Wigner transformation~\cite{Fradkin1980}
\begin{equation}
\chi_j = \left[\prod_{i=1}^{j-1} \sigma_i\right] \sigma_j\text{ and }\psi_j = \chi_j\tau_j\;\;\forall\, j~~.
\end{equation}
The parafermion operators satisfy the algebra
\begin{equation}
\chi^3=\psi^3=1;\;\chi^\dagger=\chi^2;\;\psi^\dagger=\psi^2;\; \psi\chi\psi^\dagger\chi^\dagger = \omega
\end{equation}
at each site, and
\begin{equation}
\chi_j\chi_k\chi^\dagger_j\chi^\dagger_k = \psi_j\psi_k\psi^\dagger_j\psi^\dagger_k = \psi_j\chi_k\psi^\dagger_j\chi^\dagger_k = \omega 
\end{equation}
for two sites $j,k$ such that $j<k$. The algebra implies that the modes are strongly interacting in spite of the fact that Hamiltonian in Eq-\ref{Eq:StaticHamiltonian}, when rewritten in terms of parafermions, becomes quadratic:
\begin{equation}
H =- J \omega^2 \sum_{i=1}^{L-1} \psi^\dagger_i \chi_{i+1}  -h \sum_{i=1}^L \psi^\dagger_i \chi_i  + h.c.
\label{Eq:parafermion-Hamiltonian}
\end{equation}

At $h=0$, there are two parafermionic operators $\Psi_{\rm Left}=\chi_1$ and $\Psi_{\rm Right}=\psi_L$ which commute with the Hamiltonian Eq-\ref{Eq:parafermion-Hamiltonian}. These operators cyclically change the $\mathbb{Z}_3$ charge of the states due to the commutation relation $\Psi Q=\omega Q \Psi$. As a result, $\Psi$s are associated with zero-energy parafermionic edge modes. 

Perturbative arguments presented in Ref-\onlinecite{Fendley2012}, and numerical results in Ref-\onlinecite{Jermyn2014}, suggest that regimes with an exponential decay with system size of the energy-splittings coincide with regimes with parafermionic zero-modes localized at the edge. In particular for $\arg(J)$ away from $0\Mod{\pi/3}$ , we expect a finite range of $|h/J|$ in which a zero energy parafermionic mode is localized at the edge. Since degeneracies are most robust at $\arg(J)=\frac{\pi}{6} \Mod{\frac{\pi}{3}}$, we expect the zero-mode to be strongly localized at the edge in this case.  

\subsubsection{Periodically driven $\mathbb{Z}_3$ system}

The periodically driven $\mathbb{Z}_3$ model corresponds to the $n=3$ version of the driven $\mathbb{Z}_n$ model described in Eq-\ref{drivenZn}. Since Hermiticity requires that $h_{m=1}=\bar{h}_{m=2}$ and $J_{m=1}=\bar{J}_{m=2}$, the unitary operator $U$ depends only on two complex parameters which we shall denote by $h$ and $J$. Specifically,
\begin{gather}
H_J = - J \sum_{i=1}^{L-1} \sigma_{i}^\dagger\sigma_{i+1} + h.c.\text{ and } H_h = -h \sum_{i=1}^{L} \tau_{i} + h.c.\label{PeriodicZ3}
\end{gather}
where the operators $\sigma$ and $\tau$ satisfy the $n=3$ version of the operator algebra in Eq-\ref{clockoperatoralgebra}. In the following sections, we argue that the system supports $0$ and $\pm 2\pi/3$ quasi-energy parafermion edge modes. From simple algebraic properties of the unitary operator, we infer the relative location of the regions supporting the edge modes of $0$ and $\pm 2\pi/3$ modes. The assumptions and the inferences are validated  by numerical studies of finite systems presented in the latter sections.

\section{Phase diagram for edge modes at fixed $J$}
\label{General}
\begin{figure}
\includegraphics[width=\columnwidth]{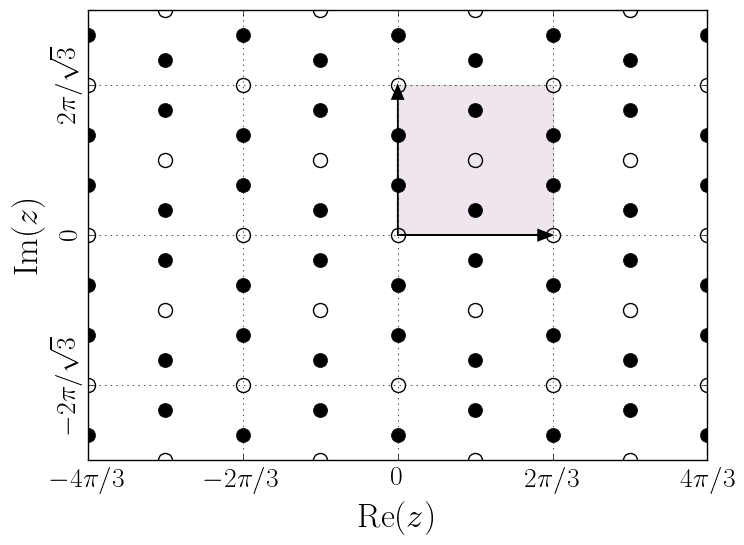}
\caption{Solutions of $z$ for $U(0,z)\propto Q^p $. Filled circles represent the case of $p=1,2$ (locations of $\frac{2\pi}{3}$ modes) and unfilled ones represent $p=0$ (locations of $0$ modes). The full lattice of solutions arises from tiling the shaded region along the arrows.\cite{Note1} }
\label{Fig:Solutions-To-z}
\end{figure}

In this section, we present general arguments predicting the qualitative structure of the edge modes in the model described in Eq-\ref{PeriodicZ3}, for a cross-section of parameter space defined by varying $h$ for a fixed $J$. 

First we make the notion of the edge mode precise in the language of a many body system. The periodic system described by $U$ carries a mode $\Psi$ of quasienergy $\epsilon$ if $\Psi$ maps any eigenstate of $U$ to a new normalizable state and satisfies $U\Psi=e^{-\imath \epsilon}\Psi U$. The mode is localized at the edge if the difference between the states $\left|\phi\right\rangle$ and $\Psi \left|\phi\right\rangle$ decays exponentially with the distance from the edge for any $\left|\phi\right\rangle$. A measure of the difference can be chosen to be the norm of the difference in the reduced density matrices away from the edge.

\subsection{Existence of non-trivial edge modes for general $\mathbb{Z}_n$}

The $\mathbb{Z}_n$-symmetric system $U(J, h=0)$ (Eq-\ref{drivenZn}) has two zero mode localized at the edges: $\Psi_{0,{\rm Left}} = \sigma_1$ and $\Psi_{0,{\rm Right}} \propto \sigma^z_L$. From the commutation relations satisfied by $\Psi$, it can be seen that they cyclically change the $\mathbb{Z}_n$ charge $Q_n$, ensuring that the action on the states are non-trivial. 

One of our central results is the connection between the existence of a zero quasienergy mode and that of modes with non-trivial quasi-energy if the parameters are shifted by an amount $z$. This connection is established, and $z$ determined, as follows. 

Assuming a zero quasienergy edge mode $\Psi_{J,h}$ (ie $\Psi_{J,h}U(J,h)=U(J,h)\Psi_{J,h}$) which cyclically changes $Q_n$ (ie $ Q_n\Psi=\omega \Psi Q_n $, $\omega$ being $n^{\rm th}$ root of unity) persists in a finite neighbourhood of $h=0$, consider the system described by $U(J,h+z)$, where $z$ is a solution to $U(0,z) \propto Q_n^p$, for a some specific integer $0\leq p<n$. 
Let $\Bra{\phi}$ be an eigenstate of the system $U(J,h+z)$ with quasienergy $E$. Action of the edge mode $\Psi_{J,h}$ on $\Bra{\phi}$ changes its quasienergy to $2\pi p/n+E$:
\begin{multline}
U(J,h+z)\Psi_{J,h}\Bra{\phi}= U(0,z)U(J,h)\Psi_{J,h}\Bra{\phi} \\= e^{\imath (\frac{2\pi p}{n}+E)} \Psi_{J,h}\Bra{\phi}
\end{multline}
where we have used $U(0,z)=Q^p_n$, $U(J,h)\Psi_{J,h}=\Psi_{J,h} U(J,h)$, and $Q_n\Psi=\omega \Psi Q_n $. 

Since the result is true for any eigenstate $\Bra{\phi}$, we have the result that the zero mode $\Psi_{J,h}$ of $U(J,h)$ is a mode of quasienergy $2\pi p/n$ of $U(J,h+z)$. 

The operator $\Psi_{J,h}$ adds quasienergies in steps of $2\pi/k$ where $k$ is the denominator of $\frac{p}{n}$ in its reduced form. In particular, for $n$ prime, for any $p\neq 0$, the edge mode adds quasienergies in steps of $2\pi/n$. For the case of $p=0$, $\Psi_{J,h}$ is a zero mode of $U(J,h+z)$.

\subsection{Algebraic determination of edge-mode location for the $\mathbb{Z}_3$ chain}

We now focus on the specific case of $n=3$. Occurrence of a mode of quasienergy $4\pi/3$ is equivalent to occurrence of one with quasienergy $2\pi/3$ as this is simply a relabeling $\Psi$ with $\Psi^2$. In order to understand the locations in the parameter space of the $2\pi/3$ modes and zero modes, we look for the solutions of the equation $U(0,z) \propto Q_3^p$,for  $p=1,2$ and $p=0$ respectively ie
\begin{equation}
\prod_{i=1}^L \exp\left(\imath z\tau_i+\bar{z}\tau_i^\dagger\right) \propto\prod_{i=1}^L \tau_i^p
\end{equation}
This is satisfied only if $\exp\left(\imath z\tau+\bar{z}\tau^\dagger\right) \propto \tau^p$. Noting that $\tau^\dagger=\tau^2$, both sides can be simultaneously diagonalized to arrive at (Appendix-\ref{App})
\begin{eqnarray}
\frac{\exp\left[2\imath {\rm Re}(z\omega) \right]}{\exp\left[2\imath {\rm Re}(z) \right]} = \omega^p \nonumber \\ 
\frac{\exp\left[2\imath {\rm Re}(z\bar{\omega}) \right]}{\exp\left[2\imath {\rm Re}(z) \right]} = \bar{\omega}^p
\end{eqnarray}
Solutions $z_0$ for the case of $p=0$, corresponding to the locations with zero mode are
\begin{multline}
z_0=\frac{\pi}{3}n + \imath \frac{\pi}{\sqrt{3}}m\\ \text{for integers }m,n \text{ such that } n+m \text{ is even}.
\label{Eq:LocationsOfZeroModes}
\end{multline}
The solutions $z$ for the case of $p=1,2$, corresponding to locations of the $2\pi/3$ modes are
\begin{equation}
z=\imath\left(1+\frac{(-1)^p}{3}\right)\frac{\pi}{\sqrt{3}} +  z_0
\label{Eq:Solutions-To-z}
\end{equation}

In summary, the arguments presented in this section imply that, since $h=0$ has an exact zero mode, the points $h=z_0$ and $h=z$ (Eq-\ref{Eq:LocationsOfZeroModes} and Eq-\ref{Eq:Solutions-To-z}), have exact zero and $2\pi/3$ modes respectively.
Additionally, if there is a domain around $h=0$ that supports a zero mode, there are domains around $h=z$ and $h=z_0$, which also support zero or $2\pi/3$ modes. All such domains have identical shape in the parameter space, since they can be translated to each other by $z$ and $z_0$.

In order to make further progress in understanding the phase diagram of the driven parafermion chain, as well as validate the assumptions made above, we rely on computational results described in the remaining sections.

\section{Numerical Results}
\label{Numerics}
\begin{figure}[ht]
\includegraphics[width=\columnwidth]{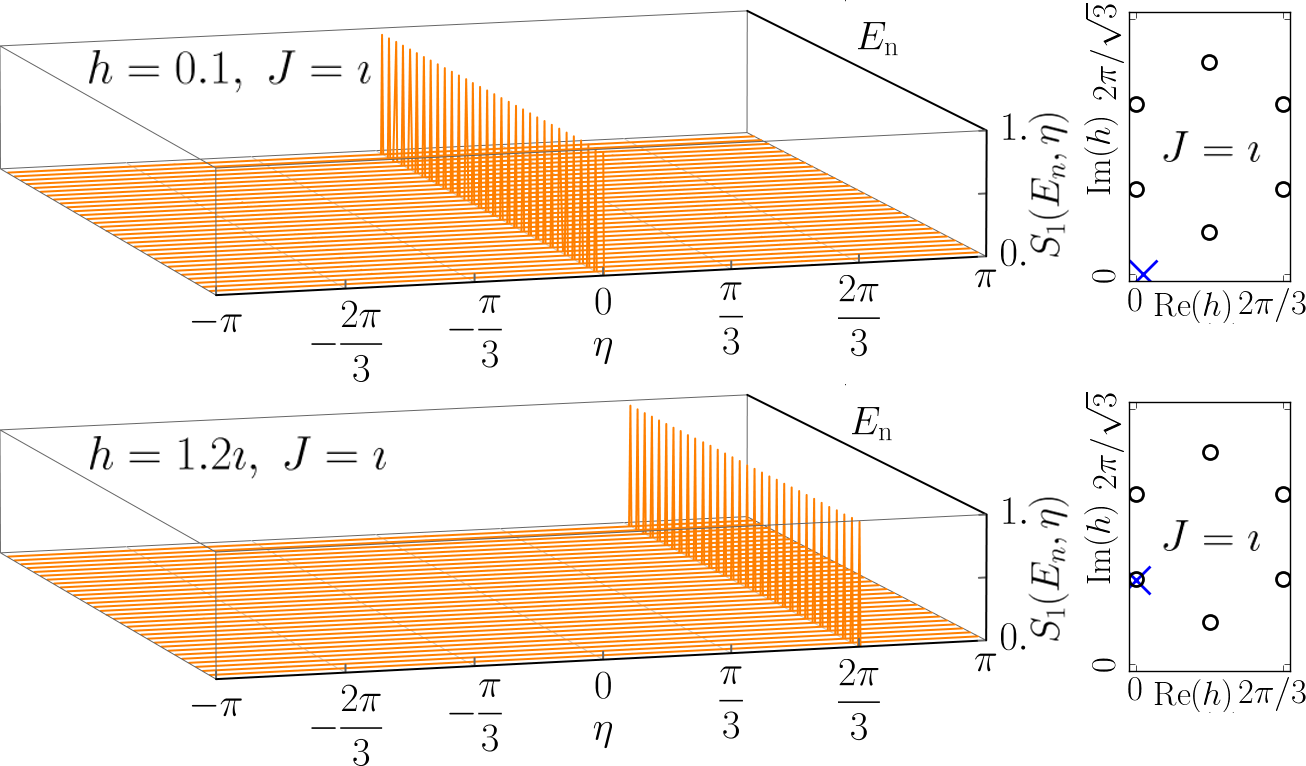}
\caption{Waterfall plots showing $S_1(E_n,\eta)$ as a function of $\eta$ for different `in-states' $E_n$ picked uniformly from the spectrum. $\eta$ corresponds to the change in quasienergy after scattering by $\sigma_1$. $S_1(E_n,\eta)$ was averaged over $50$ disorder realizations in a system of size $L=7$. Top and bottom panels show results for parameter points $(J=\imath,h=0.1)$ and  $(J=\imath,h=1.2\imath)$ respectively. Sharp peaks of strength $1$ at $\eta=0$ and $\frac{2\pi}{3}$ indicate a mode localized at the edge.  
Right panels show the complex $h$ plane in the parameter space defined by $J=\imath$, showing the solutions of $z$ (empty circles) given in Eq-\ref{Eq:Solutions-To-z}, and location of the parameters (blue cross) corresponding to the left panels.}
\label{Fig:WaterFallPlots}
\end{figure}

Our numerical studies focus on the following questions
\begin{enumerate}
	\item Does a zero mode exist in a finite region in the parameter space around $h=0$ ?
	\item How does the size of the above region depend on $\arg(J)$? Based on the results in the static case,\cite{Jermyn2014} we expect no such region for $\arg(J)=0\Mod{\frac{\pi}{3}}$, and a robust zero-mode localization for $\arg(J)=\frac{\pi}{6}\Mod{\frac{\pi}{3}}$.
	\item How does the phase diagram in the complex $h$ plane depend on $|J|$ ?
	\item Are the edge modes robust in the presence of disorder ?
\end{enumerate}
We infer the answers to the above questions as far as possible using small finite size systems by analyzing the spectral function of the clock-operator $\sigma_1$, which we describe in the following subsection.

\subsection{Edge spectral function}
\label{Sec:SpectralFunction}

\begin{figure}[ht]
\includegraphics[width=\columnwidth]{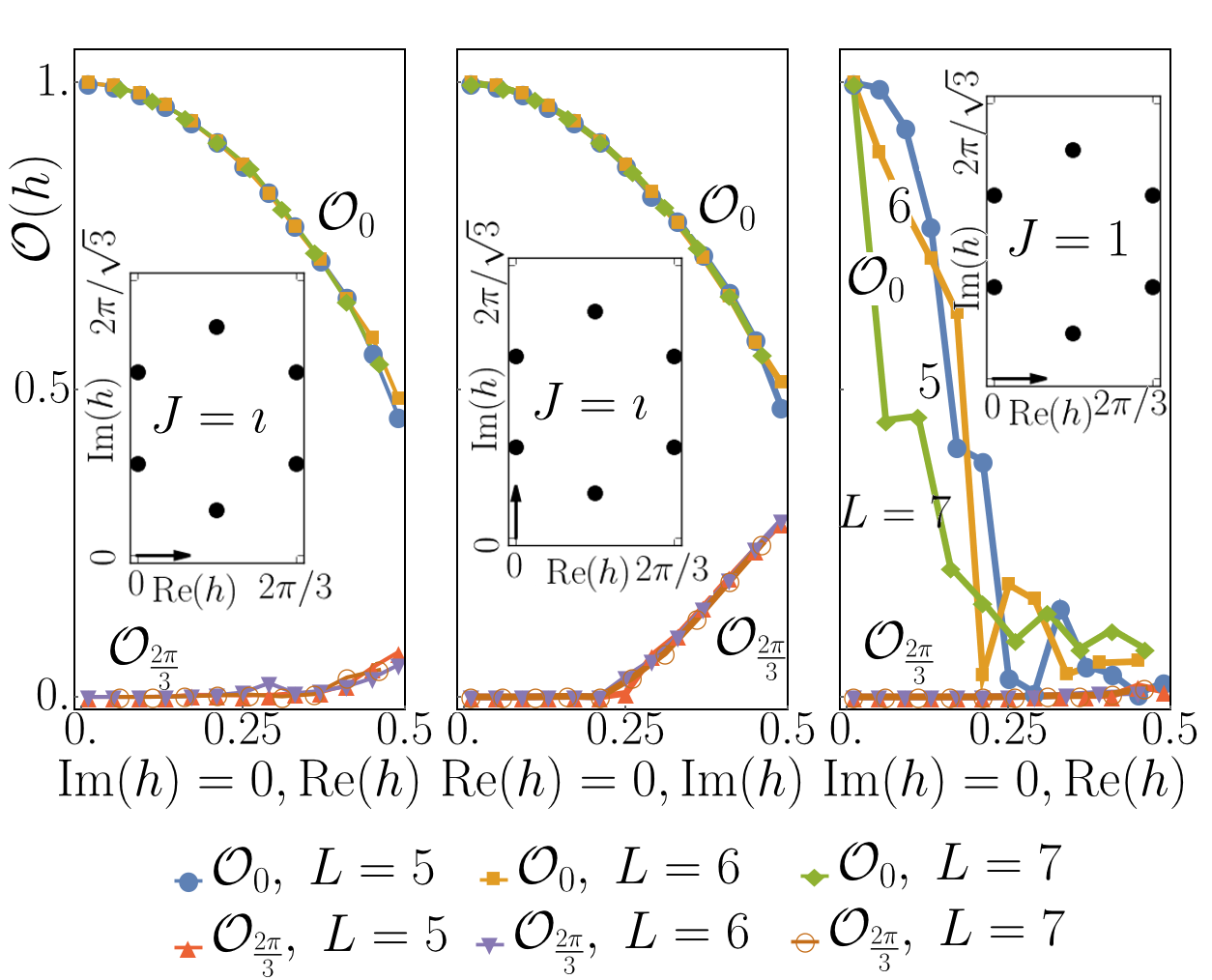}
\caption{Variation of the eigenstate order paramters $\mathcal{O}_0$ and $\mathcal{O}_{\frac{2\pi}{3}}$ defined in Eq-[\ref{EigenStateOP0},\ref{EigenStateOP2p3}] as $h$ is varied along different paths in complex parameter space depicted in the insets. left and middle panels correspond to two paths along the real and imaginary axis of $h$ in the plane defined by $J=\imath$. Right panel corresponds to $h$ varying along the real axis in the plane defined by $J=1$. Filled dots in the insets indicate solutions of $z$. Strong size dependence as well as rapid decay with away from $|h|$ of the order parameters in the case of $J=1$ (right) is indicative of the absence of a well-localised zero mode for $\arg(J)=0$.}
\label{Fig:ZeroModeAsAFunctionOfComplexhAtJ1}
\end{figure}

\begin{figure*}[!htbp]
\includegraphics[width=\textwidth]{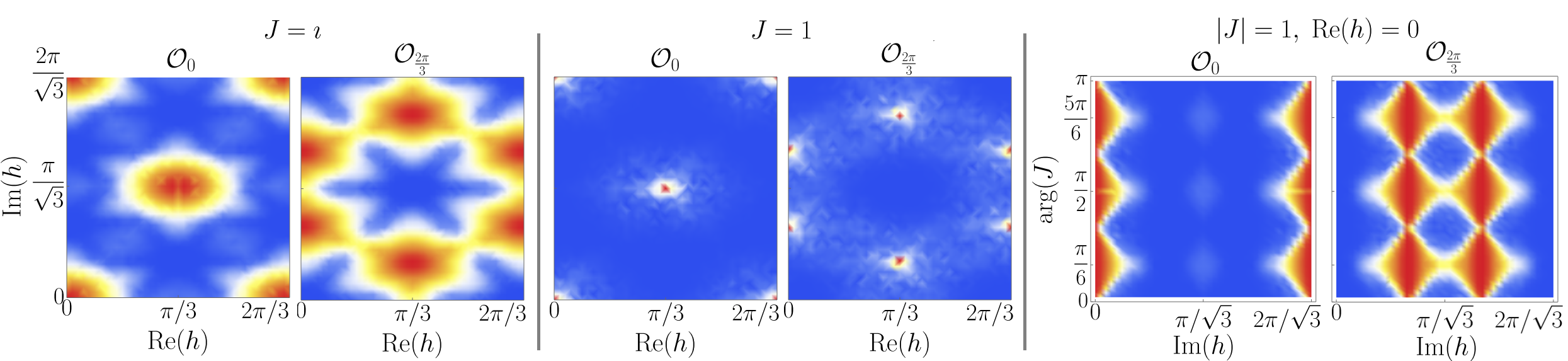}
\caption{Contour plots of $\mathcal{O}_0$ and $\mathcal{O}_{\nicefrac{2\pi}{3}}$ in the two dimensional cross sectional surfaces of parameter space defined by $J=\imath$ (left panels), $J=1$ (middle panels), and $(|J|=1, {\rm Re}(h)=0)$ (right panels). In all figures, red and blue shades represent $\mathcal{O}=1$ and $0$, respectively. All results are for systems of size $L=6$. The domains supporting edge modes are the largest and smallest when $\arg(J)=\pi/6\Mod{\pi/3}$ and $0\Mod{\pi/3}$ respectively. Fig-\ref{3D-angledependence} shows the relative locations of these planes in a three-dimensional context.}
\label{2D-angledependence}
\end{figure*}

\begin{figure}[ht]
\includegraphics[width=\columnwidth]{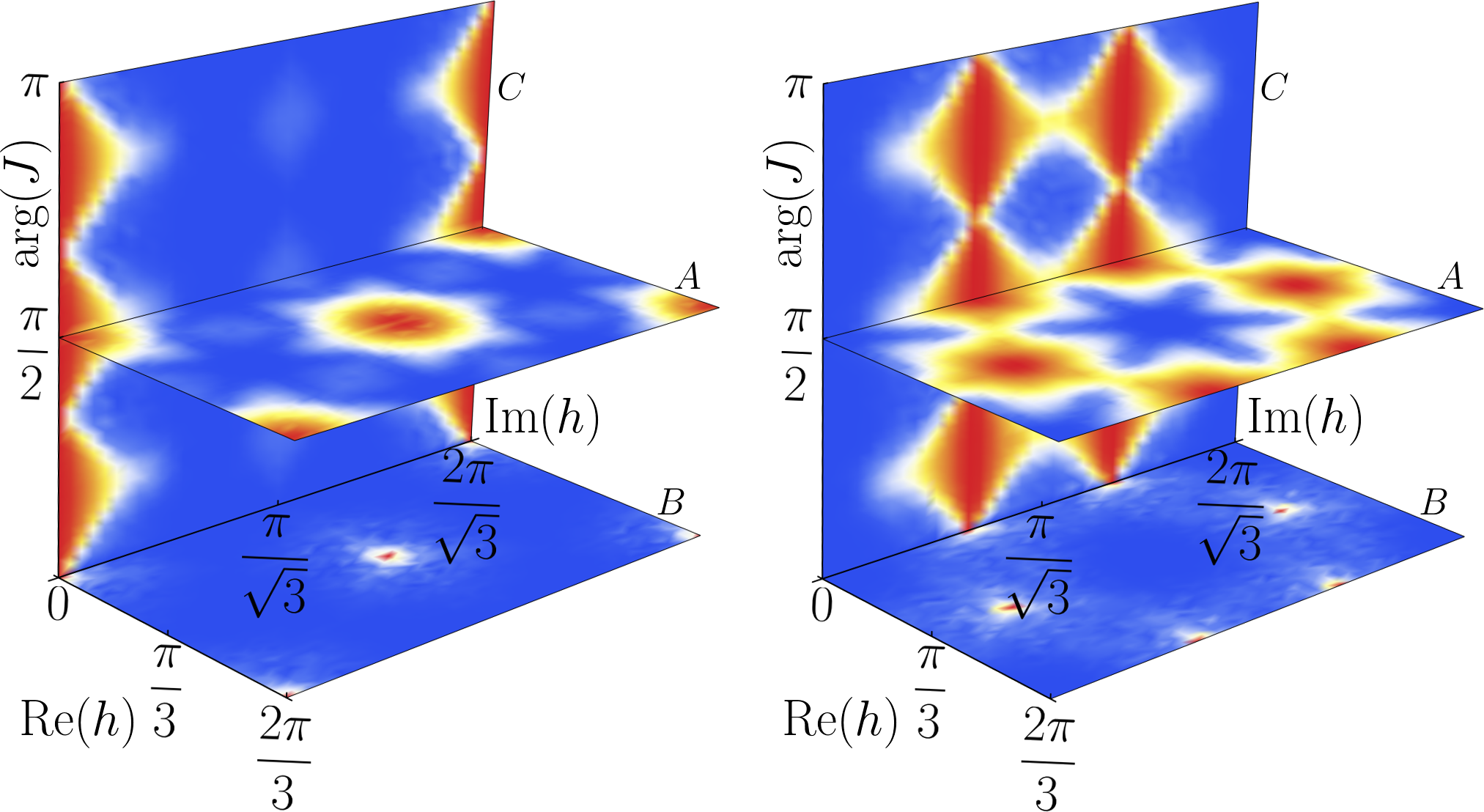}
\caption{The three dimensional cross-section of parameter space defined by fixing $|J|=1$. The planes labeled $A$, $B$ and $C$ correspond to the two dimensional cross-sections shown in the Fig-\ref{2D-angledependence} left, middle and right, respectively. Left and right panels here show $\mathcal{O}_0$ and $\mathcal{O}_{\nicefrac{2\pi}{3}}$, respectively.}
\label{3D-angledependence}
\end{figure}

The system described by $U(J,h=0)$ has a zero energy edge mode given by $\Psi_{J,h=0} = \chi_1=\sigma_1$: for any eigenstate $\Bra{E}$ of $U$,  $\sigma_1\Bra{E}$ is a linear combination of states at the same quasienergy. This is reflected as a sharp peak at $\eta=0$ in the spectral function
\begin{eqnarray}
S_1(E_n ,\eta) = \sum_{\Bra{E}} \delta (E_n-E-\eta) \left|    \left\langle E_n \left|\sigma_1\right|  E \right\rangle  \right|^2\label{Eq:spectralfunction}
\end{eqnarray}
for any state $\Bra{E_n}$ in the spectrum, $\Bra{E}$ being the eigenvectors (illustrated in system-A of Fig-\ref{Fig:WaterFallPlots}).

At finite $h$, if there is a zero energy edge mode, then a similar sharp peak should be visible in the spectral function $S_1(E_n,\eta)$ but with $\sigma_1$ replaced by $\Psi_{J,h}$. Since an exact expression for $\Psi_{J,h}$ is not available, we continue to use the spectral function $S_1(E_n,\eta)$ as defined in Eq-\ref{Eq:spectralfunction} to probe the edge mode even at finite $|h|$. As $|h|$ is increased, the spectral peak reduces in size since $\sigma_1$ now deviates from the edge mode $\Psi_{J,h}$. However if $\Psi_{J,h}$ is still localized close to the edge, this $|h|$-dependence of $S_1$ should be independent of system size. In contrast, once the localization length is of the order of the available system size, we expect strong system size dependence in $S_1(h)$, due to the lifting of the degeneracies, which should presage an absence of the edge mode. Thus system size dependence of spectral function indicates a mode that is not well-localized relative to the numerically tractable system sizes. 

In order to numerically probe the presence of a zero mode, we consider the eigenstate order parameters $\mathcal{O}_0$ and $\mathcal{O}_{\frac{2\pi}{3}}$, which describe an ordering in the many-body spectra, and which are defined as follows. $\mathcal{O}_0$ is the total spectral weight in a window of size $q=0.05$ in $\eta$, averaged over several $\Bra{E_n}$ selected from different parts of the spectra:
\begin{eqnarray}
\mathcal{O}_0 = \int_{-q/2}^{q/2}  s_1(\eta)\, d\eta
\label{EigenStateOP0}
\end{eqnarray}
where 
\begin{equation}
s_1(\eta) = \mathcal{N} \sum_{E_n\in X} S_1(E_n,\eta).
\end{equation}
$\mathcal{N}$ imposes the normalization $\int s_1(\eta) d\eta=1$, and $X$ is a set of eight eigenvectors of $U$ picked roughly uniformly from the spectrum.

To probe the presence of a $2\pi/3$-mode, we analogously use 
\begin{equation}
\mathcal{O}_{\frac{2\pi}{3}} = \int_{-q/2}^{q/2}  s_1(\frac{2\pi}{3}+\eta) + s_1(-\frac{2\pi}{3}+\eta)\, d\eta.
\label{EigenStateOP2p3}
\end{equation}
$\mathcal{O}_{\frac{2\pi}{3}}$ is defined considering the fact that in the presence of a $2\pi/3$-mode, $\sigma_1$ could in general be a linear combination of $\Psi$ and $\Psi^\dagger$ which scatters a state at $E_n$ to a linear combination of the ones at $E_n\pm \frac{2\pi}{3}$.

Fig-\ref{Fig:ZeroModeAsAFunctionOfComplexhAtJ1} illustrates the dependence of $\mathcal{O}_0$ and $\mathcal{O}_{2\pi/3}$ as $h$ is increased away from $0$ along paths in two different planes (depicted in the inset) defined by $J=\imath$ (left, middle panels) and $J=1$ (right panel). 

The rapid decay of $\mathcal{O}_0$ with $|h|$, along with the strong system size dependence in the $J=1$ case (right panel), is indicative of an absence of a localized edge-mode in the thermodynamic limit. By contrast, there are locations (middle panel) where
both order parameters appear well-converged and nonzero, indicating coexistence of both zero and $2\pi/3$ edge modes.

\subsection{Phase Diagram}

\begin{figure*}[!htbp]
\includegraphics[width=\textwidth]{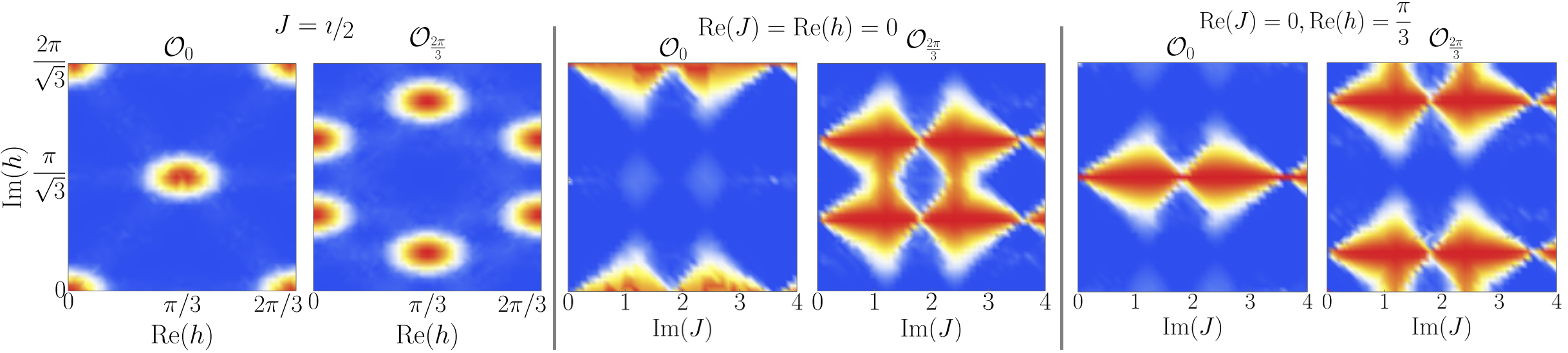}
\caption{Contour plots of $\mathcal{O}_0$ and $\mathcal{O}_{\nicefrac{2\pi}{3}}$ in the cross-sections of parameter space defined by $J={\imath/2}$(left panels). ${\rm Re}(h)={\rm Re}(J)=0$ (middle panels), and $({\rm Re}(h)=\pi/3,{\rm Re}(J)=0)$ (right panels). All results are for systems of size $L=6$. The figures show that the domains increase in size as $|J|$ increases from $0$. For larger $|J|$, the domain size varies periodically.}
\label{Fig:AbsJdependence}
\end{figure*}

\begin{figure}[ht]
\includegraphics[width=0.9\columnwidth]{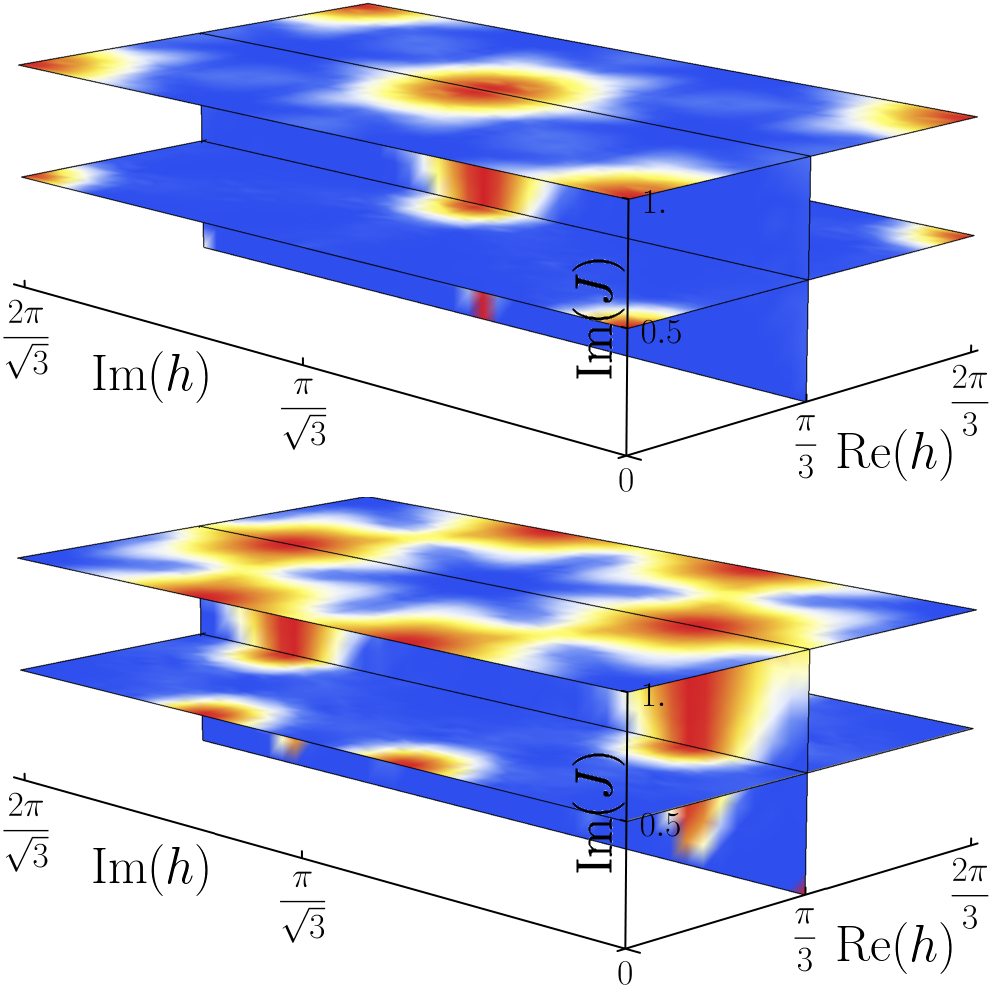}
\caption{The figures show the three dimensional section defined by $\arg(J)=\pi/2$. The planes labeled $a,b$ and $c$ show the two dimensional sections in Fig-\ref{2D-angledependence}(left), Fig-\ref{Fig:AbsJdependence}(left,right) respectively.}\label{Fig:3DargJpiBy2}
\end{figure}

In this section, we present the results from numerically scanning through various cross-sections of the parameter space. The contour plots of $\mathcal{O}$ presented here should be interpreted as phase diagrams, in the sense that regions of well-converged and typically not too small  $\mathcal{O}$ represent phases with an edge mode, while other regions do not exhibit one. Precise location of the transition lines
between these regions are not reliably pinned  down due to the small sizes accessible to our numerics. 

For different constant $J$ cross sections that we have studied, the phase diagram has the structure displayed in Fig-\ref{Fig:Solutions-To-z}. There is a parameter domain supporting a zero and a $2\pi/3$ edge mode surrounding the set of points $z_0$ (Eq-\ref{Eq:LocationsOfZeroModes}) and $z$ (Eq-\ref{Eq:Solutions-To-z}) respectively (unfilled and filled circles in Fig-\ref{Fig:Solutions-To-z}). 

Upon changing $J$, the domains supporting an edge-mode change in size but are always centered around $z$ and $z_0$, as can be seen from Fig-\ref{2D-angledependence} and Fig-\ref{Fig:AbsJdependence}. All such domains in a complex $h$-plane expand and shrink synchronously as $J$ is tuned. This is consistent with results in Sec-\ref{General}, which suggests that the domains in a single $h$-plane are images of each other under translations by $z$ and $z_0$. The size of the domains increase as $|J|$ is increased from $0$ (Fig-\ref{Fig:3DargJpiBy2}). The domains are largest and smallest for $\arg(J)=\pi/6\Mod{\pi/3}$ and $0\Mod{\pi/3}$ respectively (Fig-\ref{3D-angledependence}), similar to what was found in the static case (Sec-\ref{Sec:Z3modelStatic}).

\subsection{Effect of disorder}

\begin{figure}
\includegraphics[width=\columnwidth]{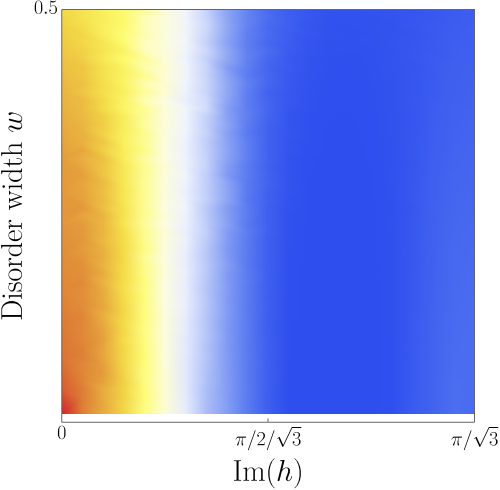}
\caption{Contour plot showing disorder averaged $\mathcal{O}_0$, for parameters along the line connecting $(J,h)=(\imath,0)$ and $(J,h)=(\imath,\imath\pi/\sqrt{3})$.}
\label{Disorderdependence}
\end{figure}

In order to probe the robustness of the edge-mode in the presence of disorder, we studied systems with disorder in the transverse field ie 
\begin{equation}
H_h = \sum_{i=1}^L h_i \tau_i + h.c.
\end{equation}
Disorder was introduced in the magnitude of the transverse field ie $h_i=(|h|_{\rm mean} + \delta h_i)\exp(\imath\arg(h_{\rm mean}))$, where $\delta h_i$ is picked from a uniform distribution over $[-w,w]$, $w$ parametrising the strength of the disorder. 
Fig-\ref{Disorderdependence} shows $\mathcal{O}_0$, averaged over 100 disorder realizations. $x$-axis corresponds to a line parametrized by ${\rm Im}(h_{\rm mean})$ between the center of a domain $(J,h)=(\imath,0)$ to $(\imath,\imath\pi/\sqrt{3})$. For the case considered here, the peak in the spectral function, appears to persist to finite disroder strength, indicating robustness of the edge mode to this form of disorder. 

Since results in Sec-\ref{General} imply that $\mathcal{O}_0(J,\{h_i\})=\mathcal{O}_\frac{2\pi}{3}(J,\{h_i+z\})=\mathcal{O}_0(J,\{h_i+z_0\})$, $z,z_0$ translate the domains supporting edge modes in a disorder averaged system into each other. The results shown here are thus representative of every domain for a given $J$.

\section{Conclusion}

In summary, we have analyzed a strongly-interacting, periodically-driven parafermion chain. We have shown numerically that there exist domains in parameter space that support zero energy edge modes. From the algebraic properties of the edge mode together with numerical methods, we have shown that the system can support $2\pi/3$ energy edge modes for $n=3$ and identified  the qualitative structure of the of the phase diagram in the four dimensional parameter space. The arguments provided here generalize to the case of $\mathbb{Z}_n$ symmetric chains, which can in general support quasienergy $\frac{2\pi}{k}$ ($k$ being a factor of $n$) modes in various parts of the phase diagram. We derive the equations governing the location of the edge modes in the appendix. 

These chains thus provide a second, intrinsically strongly interacting, example of non-trivial Floquet edge modes beyond the previously known
 $\mathbb{Z}_2$ with their $\pi$-Majorana modes transparently arising in a weakly interacting limit absent for the parafermions. Besides 
providing a concrete generalisation of Floquet edge mode physics, it can also be seen as making connection between the known and special 
$\mathbb{Z}_2$ case and more abstract classification schemes~\cite{vonKeyserlingk:2016vq,Potter:2016tb,Else:2016tj}
. 
An interesting result found in the previously-studied $\mathbb{Z}_2$ chains is a disorder-induced stabilization of a spin glass Edwards-Anderson order in higher energy eigenstates.\cite{Huse:2013bw,Khemani:2015wn}  It would be of interest to explore this physics in the clock-chain as well; however, for the system sizes currently easily accessible  a reliable finite size scaling analysis of such an order parameter is not straightforward.

\section{Acknowledgements} A.L. and R.M. thank Arnab Das, Vedika Khemani and Shivaji Sondhi for collaboration on related work. 
This work was in part supported by DFG grant SFB 1143. 

\appendix
\section{Solutions to \texorpdfstring{$z$}{z}}
\label{App}
In order to locate the domains supporting edge modes, as described in Sec-\ref{General}, we seek the solutions to $z$ of $U(0,z)\propto Q_3^p $. Substituting the definitions of $U$ (Eq-\ref{PeriodicZ3}) and $Q_3$ (Eq-\ref{ZnCharge}), we get
\begin{eqnarray}
V=\prod_{i=1}^L \exp\left[\imath (z \tau_i+\bar{z}\tau_i^\dagger)\right]&\propto &\prod_{i=1}^{n} \tau_i^p \text{ for }p=1\text{ or }2\label{Eq:ToBeSolved}
\end{eqnarray}
From tracing out all but a single lattice site, we find that this is satisfied if and only if $\exp\left[\imath (z \tau+\bar{z}\tau^\dagger)\right]\propto\tau^{p}$. Since $\tau^\dagger=\tau^2$, both sides can be simultaneously diagonalized, to obtain 
\begin{multline}
\left[\begin{array}{ccc}
\exp\left(\imath z+\imath\bar{z}\right) & 0 & 0\\
0 & \exp\left(\imath z\omega+\imath\bar{z}\overline{\omega}\right) & 0\\
0 & 0 & \exp\left(\imath z\overline{\omega}+\imath\bar{z}\omega\right)
\end{array}\right]\\ \propto\left[\begin{array}{ccc}
1 & 0 & 0\\
0 & \omega^{p} & 0\\
0 & 0 & \overline{\omega}^{p}
\end{array}\right]\nonumber
\end{multline}
which is equivalent to 
\begin{eqnarray}
\frac{\exp\left[2\imath {\rm Re}(z\omega) \right]}{\exp\left[2\imath {\rm Re}(z) \right]} = \omega^p \nonumber \\ 
\frac{\exp\left[2\imath {\rm Re}(z\bar{\omega}) \right]}{\exp\left[2\imath {\rm Re}(z) \right]} = \bar{\omega}^p \label{2p3modecondition}
\end{eqnarray}
After taking logarithms on both sides, these reduce to two linear equations:
\begin{eqnarray}
2x\left(\cos\frac{2\pi}{3}-1\right)+\imath\sin\frac{2\pi}{3} &=&	2\pi n+\frac{2\pi p}{3}\\
2x\left(\cos\frac{2\pi}{3}-1\right)-\imath\sin\frac{2\pi}{3}	 &=& 	2\pi m-\frac{2\pi p}{3}
\end{eqnarray}
where $m,n$ are any two integers and $x,y$ represent real and imaginary parts of $z$. These can be solved to obtain the solutions shown in Eq-\ref{Eq:LocationsOfZeroModes} and Eq-\ref{Eq:Solutions-To-z}, which are plotting in Fig-\ref{Fig:Solutions-To-z}. \footnote{\label{smallerunitcell}A smaller unit-cell in Fig-\ref{Fig:Solutions-To-z} would be the hexagon formed by connecting the filled dots. The whole lattice is a hexagonal lattice of such unit-cells.}

The general $\mathbb{Z}_n$ model described by Eq-\ref{drivenZn} has complex parameters $J_m$ and $h_m$ for $1\leq m\leq n-1$, satisfying $J_m=\bar{J}_{n-m}$ and $h_m=\bar{h}_{n-m}$. The condition governing the location $\vec{z}_p$ (a vector representing $\{h_m\}_{m=1\to n-1}$ and satisfying $[z_p]_m=[z_p]_{n-m}$) of $2\pi p/n$  modes ($0\leq p<n$) are the $n-1$ equations 
\begin{equation}
\frac{\exp\left[\imath \sum_m{[z_p]_m}\omega^{mq} \right]}{\exp\left[\imath \sum_m [z_p]_m \right]} = \omega^{qp},\; q=1,2,...,n-1
\end{equation}
with $\omega=\exp[2\pi\imath/n]$. Upon taking logarithms, these form a set of linear equations, with an $n-1$ dimensional lattice of solutions. For each $p$, given one solution, a whole lattice of solutions can be obtained by translating any one solution by $\vec{z}_0$. 

The edge mode $\Psi$ of $\mathbb{Z}_n$ generate a cyclic group of order $n$, and the quasienergies added by these operators form a group homomorphic to it. For $0\leq p,p'<n$, the quasienergies added by the edge operators at $\vec{z}_p,\vec{z}_{p'}$ are isomorphic if ${\rm g.c.d}(n,p)={\rm g.c.d}(n,p')$. In particular for prime $n$, edge mode quasienergies at $\vec{z}_p$ form a cyclic group of order $n$ for any $p\neq 0$, and therefore have quasienergy $2\pi/n$. For $n=6$, $\vec{z}_1$ and $\vec{z}_5$ have edge mode quasienergies forming a cyclic group of order $6$, generated by $2\pi/6$; $\vec{z}_2$ and $\vec{z}_4$ have edge modes quasienergies of order $3$ generated by $2\pi/3$; $\vec{z}_3$ has an edge mode quasienergies of order $2$ resulting in $\pi$ modes.

\bibliography{biblio_clock.bib}

\end{document}